\documentclass[aps,prl,twocolumn,showpacs,preprintnumbers,amsmath,superscriptaddress]{revtex4}
\usepackage{bm}
\usepackage{amsmath}
\usepackage[dvips]{graphicx}
\begin{document}
\bibliographystyle{apsrev}

\title{The Magnetism of  Li doped La$_{2}$CuO$_4$:
the antiferromagnetic spin-shard state}

\author{O. P. Sushkov}
\affiliation{School of Physics, University of New South Wales, Sydney 2052, Australia}
\author{A. H. Castro Neto}
\affiliation{Department of Physics, Boston University, 590 Commonwealth Avenue, Boston, 
Massachusetts 02215, USA}

\date{\today}
\begin{abstract}
We study the dynamics of a single hole in Li and Sr doped La$_{2}$CuO$_4$ 
and its extension to a finite hole concentration.
We compare the physics of La$_{2-x}$Sr$_x$CuO$_4$ and 
La$_{2}$Cu$_{1-x}$Li$_x$O$_4$ and explain why these systems are 
remarkably different.
We demonstrate that holes in La$_{2}$Cu$_{1-x}$Li$_x$O$_4$ are always localized and
that there is a critical concentration, $x_c\approx 0.03$, above which 
the holes break the global antiferromagnetic 
state into an array of weakly coupled antiferromagnetic clusters 
(antiferromagnetic shards). We show that the spin-shard state provides a
description of the magnetic and electric properties of 
La$_{2}$Cu$_{1-x}$Li$_x$O$_4$. Two experiments that can test our theory are proposed.
\end{abstract}
\maketitle

{\it Introduction.} The sharp contrast in the physical properties of Sr and Zn
doping in La$_{2}$CuO$_4$ is not surprising since Sr brings mobile holes into 
the CuO$_2$ planes while Zn does not. At the same time the remarkable 
differences between Li and Sr doping of La$_{2}$CuO$_4$ are somewhat 
surprising since both Sr and Li ions bring mobile holes into CuO$_2$ planes.
It is by now well established that La$_{2-x}$Sr$_x$CuO$_4$ (LSCO) and
La$_{2}$Cu$_{1-x}$Li$_x$O$_4$ (LCLiO) have very different physical properties:
LSCO has an insulator-superconductor transition 
at $x \approx 0.055$ \cite{Kastner} 
while 
LCLiO remains an insulator at all dopings \cite{Sarrao}.
Elastic and inelastic neutron scattering in 
LSCO at $x>0.02$ reveals incommensurate magnetic peaks
\cite{Yamada,Wakimoto,Matsuda,Fujita} while neutron scattering in
LCLiO is always commensurate \cite{Bao,Chen1,Chen2} with a very unusual scaling
behavior of the 
magnetic response \cite{Chen1,Chen2}.
There is, however, a similarity between these compounds: the long-range
N\'eel order is destroyed at rather close values of doping, $x =0.02$ in
LSCO and $x =0.03$ in 
LCLiO\cite{Kastner,Heffner}.

It is clear that the origin of the differences between Sr and Li doping is in
the fact that 
Sr substitutes La ions outside of the CuO$_2$ planes while Li ions substitute
in plane for Cu ions. Every dopant injects a hole into the plane, but the Sr
ion seems to be a relatively weak hole scatterer while the Li ion interacts
strongly with the injected holes.
To understand the differences between these two systems we first consider 
the problem of an impurity bound state formation in a strongly correlated
CuO$_2$ plane. 
Then we generalize our solution for a finite concentration
of holes and find that there is a critical concentration $x_c$ above
which a spin-shard state is formed. We show that the formation of this state
can explain the observed properties of 
LCLiO.

{\it Coulomb trapping of a single hole.}  The problem of a single hole
trapped by a Sr ion has been solved recently in Ref. \cite{SK1}. In this work
we extend that solution to the case of Li doping.  Similarly to the case
studied in Ref. \cite{SK1} we perform the analysis within the $t-t'-t''-J$ model 
that is assumed to describe the low energy, $\epsilon \leq 4t \sim 1eV$, dynamics of
CuO$_2$ planes. The parameters of the effective  $t-t'-t''-J$ model 
are well known: from the Raman scattering studies \cite{Tokura} it was found 
that $J \approx 125$ meV; following the calculations of Andersen {\it et al.} \cite{And} 
we set $t/J =3.1$, $t'/J \approx -0.5$, and $t''/J \approx 0.3$.
We measure the energy in units of $J$, length in units of the
lattice spacing $a$, and set $\hbar =1=k_B$. 
The kinetic terms $t'$ and $t''$ are small compared to $t$ and our results
for  
LCLiO are not {\it qualitatively} sensitive to
their precise values\cite{com1}. 
Nevertheless, these matrix elements are quantitatively important because they describe hopping
within the same magnetic sublattice and influence the hole dispersion.

Because of the strongly interacting environment where the hole moves, the
hole dynamics is strongly dressed by quantum spin fluctuations with momenta
${\bf q} \sim (\pi,\pi)$. 
This dressing leads to the effective hole (quasihole or small magnetic
polaron) that describes dynamics of the system  at even lower energies, $\epsilon \le J$. 
This part of the problem is well understood and is independent of 
the presence of the impurity. In momentum space, the minima of the quasihole
dispersion appear near ${\bf k}_0=(\pm\pi/2,\pm\pi/2)$, which are the centers
of the four faces of the reduced magnetic Brillouin zone (MBZ). 
In vicinity of these points the dispersion is quadratic:
$\epsilon_{\bf k}\approx \beta_1 k_1^2/2+ \beta_2 k_2^2/2$, 
where ${\bf k}$ is defined with respect to ${\bf k}_0$, ${\bf k}_1$ is perpendicular
to the face of the MBZ, while ${\bf k}_2$ is parallel to it. 
In the self-consistent Born approximation \cite{SK} 
the hole dispersion is isotropic, $\beta_1\approx\beta_2=\beta\approx 2.2$, 
and has a finite quasiparticle residue, $Z \approx 0.34$.

In the presence of a negatively charged impurity, the hole is trapped near
the impurity ion by the Coulomb potential $-e^2/(\epsilon_{e} \sqrt{r^2+d^2})$,
where $\epsilon_{e}\sim 30 - 100$ is the effective dielectric constant 
of  La$_{2}$CuO$_4$ \cite{Kastner}, and $d$ is the distance of the ion 
from the CuO$_2$ plane. In the cases discussed here we can set $d=0$ 
since its effect is rather small. The most important difference between the Li and the
Sr doping is that the hole cannot penetrate on the Li site
because of a strong hard core repulsion. Thus, we can model
the potentials for Sr and Li as:
\begin{eqnarray}
{\rm Sr}: \  V(r)=-\frac{e^2}{\epsilon_{e}r}  , 
\ \ \ {\rm Li}: \ V(r)=-\frac{e^2}{\epsilon_{e}r}+G\delta({\bf r}) \, , 
\nonumber
\end{eqnarray} 
where $G$ is a very large positive constant. 
The $\delta$-function term in the Li potential enforces a node of the
bound-state wave function at the origin. So the ground state wave function in
the Li case can have either s-wave with a node, 
or p-wave symmetry.
 
Because of the presence of the bound state the local magnetic
configuration relaxes, leading to the formation of a spin texture:
the spin at every site can rotate by an angle $\theta_i$ with respect to the 
orientation at $r=\infty$. There are two sublattices, ``up'' and ``down'', 
so that:
\begin{eqnarray}
|\mbox{i}\rangle&=&e^{i\theta({\bf{r_i}}){{\bf{m}}}
\cdot{\bf{\sigma}}/2}|\uparrow\rangle,   \
\mbox{ \  i $\in$  ``up'' \ sublattice}  ,\nonumber\\
|\mbox{j}\rangle&=&e^{i\theta({\bf{r_j}}){\bf{m}}
\cdot{\bf{\sigma}}/2}|\downarrow\rangle,   \
\mbox{ \  j $\in$  ``down'' \ sublattice} .
\nonumber
\end{eqnarray} 
Here ${\bf m}= (\cos\alpha,\sin\alpha,0)$ with arbitrary $\alpha$  is  the 
``director'' of the state. The director is orthogonal to the magnetization plane.
Note that the directions in spin space are completely independent of directions
in  coordinate space. 
Because of the sublattice structure, the wave function of the hole $\psi({\bf
  r})$ has 
two components, pseudospin up and pseudospin down,
corresponding to a hole moving in the up and down sublattices, respectively. 
The total  energy is of the form \cite{IF,SK}:
\begin{eqnarray}
\label{Hab1} 
&&E=\int d^2r \left\{\frac{\rho_s}{2} (\nabla\theta)^2+\right.\\
&&\left. \psi^{\dag}({\bf r})\left(\begin{matrix}
-\beta\frac{\nabla^2}{2}-\frac{e^2}{\epsilon_er} \\
\sqrt{2}g e^{i\alpha} ({\bf e}\cdot{\bf \nabla}\theta)
\end{matrix}
\begin{matrix}
\sqrt{2}g e^{-i\alpha} ({\bf e}\cdot\nabla\theta)\\
-\beta\frac{\nabla^2}{2}-\frac{e^2}{\epsilon_er}
\end{matrix}\right) \psi({\bf r}) \right\} \ ,\nonumber
\end{eqnarray}
where  $\rho_s/J \approx0.18$ is the spin stiffness of environment and 
${\bf e}$  is a unit vector orthogonal to a given face of the MBZ.
If the hole ``resides'' in the pocket near ${\bf k}_0=(\pi/2,\pi/2)$ then
${\bf e}=(1/\sqrt{2},1/\sqrt{2})$, and if the hole ``resides'' in the pocket near 
${\bf k}_0=(\pi/2,-\pi/2)$ then ${\bf e}=(1/\sqrt{2},-1/\sqrt{2})$.
There is coupling between different pockets, but this is an exponentially
small effect and can be safely neglected.
Finally, $g=Zt\approx 1.05$ is the effective hole-spin-wave coupling constant.
The solution of (\ref{Hab1}) is of the form
\begin{equation}
\label{psis}
\psi({\bf r})=\frac{1}{\sqrt{2}}
\left(\begin{matrix}
1 \\
-e^{i\alpha}
\end{matrix}
\right)\chi(r) \, ,
\end{equation}
where, 
\begin{eqnarray}
\label{wfSrLi}
{\rm Sr}: \ \chi(r)= \sqrt{\frac{2}{\pi}}\kappa e^{-\kappa r} \ , 
\ \  {\rm Li}: \  \chi(r)= \sqrt{\frac{4}{3\pi}}\kappa'^2  r  e^{-\kappa' r} \, , 
\end{eqnarray}
with $\kappa$ and $\kappa'$ as
variational parameters. Note that these solutions have s-wave
symmetry. We have checked that the s-wave state is lower in energy
than the p-wave one.

A variation of the energy (\ref{Hab1}) with respect to $\theta$ leads to,  
\begin{equation}
\label{eqph}
\nabla^2 \theta =\sqrt{2} g/\rho_s ({\bf e}\cdot{\bf \nabla})\chi^2(r) \ ,
\nonumber
\end{equation}
which has the solutions:
\begin{eqnarray}
\label{solthet}
{\rm Sr}: \ \ \ \theta=&&\frac{g}{\sqrt{2}\pi\rho_s}\frac{({\bf e\cdot r})}{r^2}
\left[1-e^{-2\kappa r}(1+2\kappa r)\right] \ , \nonumber\\
{\rm Li}: \ \ \ \theta=&&\frac{g}{\sqrt{2}\pi\rho_s}\frac{({\bf e\cdot r})}{r^2}
 [1-e^{-2\kappa' r}(1+2\kappa' r\nonumber\\
&&+2\kappa'^2r^2+\frac{4}{3}\kappa'^3r^3)] \ .
\end{eqnarray}
Substitution of these solutions together with (\ref{psis}) and (\ref{wfSrLi}) in Eq.~(\ref{Hab1}),
and the minimization of energies with respect to $\kappa $ and $\kappa'$ 
leads to:
\begin{eqnarray}
\label{ks}
{\rm Sr}: && \kappa=\frac{2 e^2/(\epsilon_e \beta)}{\left(1-\frac{g^2}{2\pi\beta\rho_s}\right)}, \ 
E_{{\rm
    Sr}}=-\frac{\beta\kappa^2}{2}\left(1-\frac{g^2}{2\pi\beta\rho_s}\right) , 
\\
{\rm Li}: && \kappa'=\frac{2 e^2/(\epsilon_e \beta)}{\left(1-\frac{5g^2}{16\pi\beta\rho_s}\right)}, \ 
E_{{\rm Li}}=-\frac{\beta\kappa'^2}{6}\left(1-\frac{5g^2}{16\pi\beta\rho_s}\right) \ . \nonumber
\end{eqnarray}
The effective dielectric constant $\epsilon_e$ is not known accurately.
Therefore, to determine $\kappa$ we rely on the estimates that follow from the hopping 
conductivity \cite{Kastner,SK}: $\kappa \approx 0.4 \approx 0.1\AA^{-1}$.
Then, from (\ref{ks}), one finds that $\kappa'\approx 0.31 \approx 0.08\AA^{-1}$, 
$E_{{\rm Sr}}/J\approx -0.09$ (i.e., $E_{{\rm Sr}} \approx -11$ meV),
and $E_{{\rm Li}}/J\approx -0.023$ (i.e., $E_{{\rm Li}} \approx -3$ meV).
Although the binding energies are subject to uncertainties associated
with the value of the dielectric constant, the ratio $\kappa/\kappa'\approx 1.3$ 
and hence relative sizes of Sr and Li bound states are reliable.
According to measurements \cite{Park} of the dielectric response of
LCLiO, the charge excitation gap at $x=0.023$ is
$\Delta\approx 1$ meV. This value is in a reasonable agreement
with our estimate $|E_{{\rm Li}}|\approx 3$ meV especially having in mind
that our result is relevant only to the very low doping, $x\ll 0.03$. 
We will argue below that the gap decreases with
doping switching from one regime to another at $x\approx 0.03$.

Plots of radial charge density $2\pi r \chi^2(r)$ for Sr and Li impurities are 
given in Fig.1a.
\begin{figure}
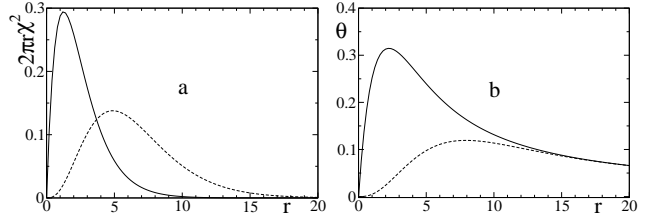

\centering
\includegraphics[height=80pt,keepaspectratio=true]{Fig1a.eps}
\includegraphics[height=80pt,keepaspectratio=true]{Fig1b.eps}
\vspace{-5pt}
\caption{{\bf a}: The radial charge density $2\pi r \chi^2$ versus radius.
{\bf b}: The angle $\theta$ (radians) of the background spin rotation versus radius. The 
angle is given for direction ${\bf r} || {\bf e}$.
The solid and the dashed lines correspond to Sr and Li impurities, respectively.}
\label{Fig1}
\end{figure}
\noindent
In Fig.1b we show the angle of the spin rotation of the antiferromagnetic
background for Sr and Li impurities (see Eqs. (\ref{solthet})). 
Inside the ``atomic core'', $r \ll 1/\kappa$, the Sr solution (\ref{solthet})
describes a uniform 
$(1,\pm1)$ spiral, $\theta \approx \sqrt{2}g \kappa^2 ({\bf e\cdot
  r})/(\pi\rho_s)$, 
while the Li solution (\ref{solthet}) gives a  very small rotation angle, 
$\theta \approx g \kappa^4r^2 ({\bf e\cdot r})/\rho_s$. This result
indicates that while Sr impurities induce incommensurate magnetic textures,
Li impurities do not lead to a significant change in the commensurate state.

{\it Destruction of the long-range order.} 
According to Eq. (\ref{solthet}), at large
distances, 
$r \gg 1/\kappa$, the Sr and Li ions generate a similar
disturbance in the antiferromagnetic background: 
$\delta {\bf n}=[{\bf n_0}\times{\bf m}]\theta ={\bf M} ({\bf e\cdot r})/(2\pi r^2)$,
where ${\bf n}=\delta {\bf n}+{\bf n_0}$ is the N\'eel unit vector,
${\bf n_0}={\bf n}(r=\infty)$, and $|{\bf M}|=\sqrt{2}g/\rho_s \approx 8$ is 
the ``dipole moment'' of the bound state.
It has been shown in Refs. \cite{Glazman,Cherepanov} that the dipoles 
frustrate the antiferromagnetism 
and hence lead to destruction of the N\'eel order at some critical 
concentration, $x_c$. 
Since $M$ is more or less the same for LSCO and LCLiO, 
we would expect $x_c$ to be the same in these two compounds. 
For $M \approx 8$ one finds $x_c\approx 0.02$ \cite{SK1}.
However, for $x\sim 0.02$ the average separation between ions
is $r\sim 7 a$ and hence, according to Fig.1b 
the effective dipole moment of the Li impurity is somewhat smaller than the one
of Sr. Thus, the change in the dipole moment leads to a value of
$x_c \approx 0.03 $ for LCLiO 
that is slightly larger than the value of LSCO ($x_c \approx 0.02$). 
This result explains the destruction of N\'eel order in LCLiO. 

In spite of similarity at very small $x$, the physics at $x > x_c$
in 
LSCO and 
LCLiO is quite different.
It has been argued in Ref. \cite{SK1} that for $0.02 < x <0.055$ LSCO
remains insulating because the bound states have not percolated. 
Nevertheless, because of  
the long-range dipole-dipole interaction between the bound states, the
insulating spin-glass-like  state with $(1,\pm 1)$ spiral ordering is 
established \cite{HCS}. Due to the Dzyaloshinskii-Moriya interaction the
spiral direction is pinned to the orthorhombic b-axis.
The size of the Sr-hole bound state shown in Fig.1a is consistent with this picture.
At $x \approx 0.055$ the Sr-hole bound states percolate and, due to the Pauli
principle, there is a rotation of the spin spiral 
direction by 45$^\circ$ to the (1,0) or (0,1) incommensurate order \cite{SK1}. 
At $x > 0.055$  the charge distribution is more homogeneous.

According to Fig.1a the size of the Li bound state is a factor 1.5 - 2 larger that that for Sr.
Therefore in 
LCLiO there is no window of doping between destruction of the N\'eel order and 
percolation of bound states. Hence, percolation occurs at $x\approx x_c\approx
0.03$. However, after the percolation the system remains extremely
inhomogeneous because of strong scattering of holes by Li ions.
The hole localization length should be of the order of the mean free path 
that is of the order of the distance between impurities.
Since the localization length is the same as the average separation between
holes, the regions occupied by a single hole and regions occupied by two
(or a few) holes are equally probable.
It has been shown in Ref. \cite{SK1} that single hole regions induce $(1,\pm
1)$ spirals while few hole regions induce (1,0) or (0,1) spirals. Thus, at
any doping above the 
percolation the system remains
completely frustrated and no collective spiral order can be established. The
system consists 
of weakly and randomly coupled
antiferromagnetic clusters. We call these clusters antiferromagnetic shards. 

This picture leads to a prediction for the nuclear quadrupole resonance (NQR)
in LCLiO.
It is known that in 
LSCO the Cu NQR A-line is shifted with respect to its position 
in the parent compound La$_2$CuO$_4$, and the shift is proportional to doping \cite{Ohsugi}.
The A-line is wide, reflecting a non-uniform charge distribution \cite{Haase}.
In the spin-shard state of 
LCLiO (i.e. at $x> 0.03$) one should expect two distinct NQR 
lines of comparable intensity (two Cu lines and two O-lines).
The first one originates from antiferromagnetic shards.
It is likely that the frequency of this line is close to that in the parent compound.
The second line originates from regions occupied by holes and, 
therefore it should have a higher frequency and larger width. We note
that inhomogeneities have already been observed in La-NQR in the ordered phase 
($x<0.025$) \cite{suh}. We assign those to the formation of bound
states discussed here. 

{\it The spin-shard state.}
Below percolation, $x <x_c$ the system retains the long-range 
antiferromagnetic order.
Above percolation, $x > x_c$ there are only finite antiferromagnetic shards 
with $N$ spins because all exchange interactions are short range.
In fact, there is a distribution of shard sizes and the probability
of finding a shard of size $N$ is given by percolation theory \cite{StAh}:
\begin{equation}
\label{PN}
P(N,x)\propto \exp\{-N/N(x)\} \ ,
\end{equation}
where $N(x)$ is the average shard size at doping $x$.
Because of magnetic anisotropies in the system (Dzyaloshinskii-Moriya and
crystal field) the low energy physics of the shard can be thought as
a two-level system with a tunneling splitting, $\Delta(N)$, given
by the WKB expression:
\begin{equation}
\label{DN}
\Delta(N)\propto \omega_0 \exp\{-\gamma N\} \ ,
\end{equation}
where $\omega_0\propto J$ is an attempting frequency and
$\gamma$ is a parameter that depends on the magnetic anisotropies.
The combination of (\ref{PN}) and (\ref{DN}) shows that there is a distribution of tunneling 
splittings in the problem that has the form:  
\begin{equation}
\label{PD}
P(\Delta)\propto \Delta^{\alpha-1} \ ,
\nonumber
\end{equation}
where 
$\alpha=1/(\gamma N(x))$ is a non-universal quantity which depends on doping. 
The neutron scattering is peaked at ${\bf q}=(\pi,\pi)$, 
and the imaginary part of the average susceptibility is given by:
\begin{equation}
\label{chi''}
\chi_{\mu}''(\omega,B,T)= \int d\Delta P(\Delta) \chi''_{\mu,{\rm shard}}(
\omega,B,T,\Delta) \ ,
\end{equation}
where $\mu = 0,1$ refers respectively to longitudinal and transverse 
polarizations of the neutron magnetic field with respect to the
applied uniform external magnetic field $B$.
The shard susceptibility is given
by the two-level system expression ($\omega>0$):
\begin{eqnarray}
\chi''_{\mu,{\rm shard}} &\propto& 
\frac{(\Delta^2)^{1-\mu} ({\cal M}^2)^{\mu} }{\Delta^2+{\cal M}^2}
\nonumber 
\\
&& \tanh\left(\frac{\sqrt{{\cal M}^2+\Delta^2}}{2T}\right)
\delta(\omega-\sqrt{{\cal M}^2+\Delta^2}) \ ,
\nonumber
\end{eqnarray}
where ${\cal M}(B) \propto \mu_B\sqrt{N} B$ is the Zeeman energy 
(since a shard of $N$ spins is a random system, 
there are on average $\sqrt{N}$ more spins in one sublattice than in the other
leading to a net shard moment).
Hence, the integration in (\ref{chi''}) gives the following result:
\begin{eqnarray}
\label{chi''1}
\chi_{\mu}''(\omega,B,T) = \frac{1}{T^{1-\alpha}} F_{\mu}(\omega/T,{\cal M}/\omega)  \ ,
\end{eqnarray}
where
\begin{eqnarray}
F_{\mu}(x,y) \propto x^{\alpha-1} y^{2 \mu} (1-y^2)^{\alpha/2-\mu} \tanh(x/2) 
\Theta(1-y)
\nonumber
\end{eqnarray} 
is a scaling function \cite{logs}, 
where $\Theta(x) = 1 (0)$ if $x>0$ ($x<0$) is the Heaviside step function.
In a unpolarized neutron scattering experiment both longitudinal and
transverse responses appear. When $y \sim 1$, as in the case of nuclear
magnetic resonance (NMR) \cite{abragam},
 the transverse response is the dominant one.
In this case, since we expect $\alpha <1$, the transverse susceptibility
($\mu = 1$) becomes singular at $y = 1$ while the longitudinal response
($\mu =0$) vanishes. This resonant effect should be
easily observable in neutron scattering experiments in a finite magnetic
field because the damping to the shard motion should be weak due to
the insulating nature of the background.  
Equation (\ref{chi''1}) is in agreement with zero magnetic field 
($\mu=0,y \to 0$)
experimental data of Ref. \cite{Chen1}
with $\alpha \approx 0.35$ for $x=0.04$ and $\alpha \approx 0.06$ for
$x=0.06$ and $x=0.1$. By performing the same experiments in a
magnetic field one can test our prediction for (\ref{chi''1}).

In reality the antiferromagnetic shards are not decoupled.
Shards from different CuO$_2$ planes interact
because of the interplane exchange $J_{\perp}$. Due to this interaction
the system must freeze to a disordered 3D spin-shard glass at the 
characteristic temperature $T_{3D}\sim J_{\perp}N_{ov}$,
where $N_{ov}\sim N(x)$ is the number of spins in two shards that
overlap along the c-axis.
Thus, Eq. (\ref{chi''1}) is relevant to the spin-shard state at $T > T_{3D}$
where shards fluctuate independently.
At $T < T_{3D}$ the shards freeze. However,  Eq. (\ref{chi''1}) is still 
valid, but instead of the current temperature $T$ one must substitute the 
freezing temperature $T_{3D}$. Thus, for $T<T_{3D}$ one finds that
$\chi''(\omega,T)$ should be roughly temperature independent, in agreement
with the experimental data \cite{Chen1,Chen2}.

In conclusion, based on an analysis of the extended $t-J$ model,
we propose the theory of antiferromagnetic-spin-shard state for
La$_{2}$Cu$_{1-x}$Li$_x$O$_4$.
The theory explains why the long-range N\'eel order is destroyed at $x_c=0.03$.
At $x<x_c$ physics of the system is driven by Li-hole bound states.
At $x=x_c$ the N\'eel order is destroyed and simultaneously the bound states percolate.
At $x>x_c$ the holes in La$_{2}$Cu$_{1-x}$Li$_x$O$_4$  remain localized breaking the
global antiferromagnetic state  into array of weakly coupled 2D antiferromagnetic shards. Due to the interlayer interaction the shard state freezes to a
glassy state (spin-shard-glass) below some characteristic temperature $T_{3D}$.
While our theory describes quite well the available experimental data, it
also predicts two independent effects: the existence of two distinct Cu(O)-NQR
lines, and a magnetic field scaling of the neutron scattering data.

We would like to thank Wei Bao, Anders Sandvik, and John Sarrao for
stimulating discussions. 
O.P.S. acknowledges the Quantum Condensed Matter Visitor's
Program at Boston University, and the Kavli Institute for Theoretical 
Physics at UCSB (support by the NSF  Grant No. PHY99-07949) for their hospitality. 
A.H.C.N. was supported through NSF grant DMR-0343790.


\begin{thebibliography}{99}
\bibitem{logs}In writing (\ref{chi''1}) we have disregarded logarithmic
corrections.
\end{thebibliography}
\end{document}